# Optimizing Adaptive Video Streaming: A Fuzzy Logic Approach to Resource Allocation

Koffka Khan[1]

[1]Department of Computing and Information Technology, Faculty of Science and Agriculture, The University of the West Indies, St. Augustine Campus, TRINIDAD AND TOBAGO.
Email address: koffka.khan@gmail.com

*Abstract*— The demand for high-quality video streaming has propelled the evolution of adaptive streaming systems. Efficient resource allocation is paramount to ensuring optimal viewer experience, considering dynamic factors such as server load, network bandwidth, and viewer demand. This review paper investigates the application of fuzzy logic to enhance resource allocation in adaptive video streaming. Fuzzy logic, known for its adaptability in uncertain environments, offers a promising approach to address the complexities of resource optimization. We delve into the integration of fuzzy logic in resource allocation models, considering key parameters like server load, network bandwidth, and viewer demand. The paper provides a comprehensive examination of the benefits, challenges, and limitations associated with fuzzy logic-based resource allocation, supported by case studies illustrating successful implementations. Evaluation metrics are discussed to gauge the performance of this approach in comparison to traditional methods. The review concludes with insights into future directions, offering a roadmap for further advancements in the field of adaptive video streaming. "Optimizing Adaptive Video Streaming: A Fuzzy Logic Approach to Resource Allocation" aims to contribute to the ongoing discourse on intelligent resource management for enhanced video streaming quality and user satisfaction.

## I. INTRODUCTION

Adaptive video streaming is a dynamic and responsive content delivery technique designed to optimize the viewing experience for users in varying network conditions. Unlike traditional streaming methods, adaptive streaming adjusts the quality of video content in real-time, ensuring seamless playback by adapting to the viewer's available bandwidth. This technology is particularly crucial in the contemporary landscape, where users consume video content on a diverse range of devices and networks. By providing different video quality levels and resolutions, adaptive streaming aims to maintain a consistent and high-quality viewing experience, irrespective of fluctuating network conditions.

Adaptive video streaming faces inherent challenges related to resource allocation, as it necessitates the efficient utilization of server resources, network bandwidth, and other crucial components. Balancing these resources becomes increasingly complex in dynamic environments where network conditions can fluctuate rapidly. The challenge lies in delivering optimal video quality while avoiding issues such as buffering, stuttering, or long loading times. Traditional streaming systems often struggle to adapt to these variations, leading to a suboptimal viewing experience. Therefore, addressing resource allocation challenges is pivotal for ensuring the continuous and high-quality delivery of video content in adaptive streaming systems.

Intelligent and dynamic resource allocation methods [35], [34], [4], [7], [2] are imperative to overcome the complexities associated with adaptive video streaming [29], [25], [32], [36], [23]. In traditional streaming systems, static resource allocation may lead to underutilization or overutilization of resources, resulting in either inefficient delivery or a poor user experience. Adaptive streaming, on the other hand, demands a more sophisticated approach. Intelligent resource allocation involves real-time assessment and adaptation to factors such as server load, network conditions, and viewer demand. The need for dynamic allocation stems from the ever-changing nature of network environments and user behavior. Implementing intelligent algorithms, such as fuzzy logic, allows for the continuous optimization of resource allocation, ensuring that the streaming system can respond dynamically to varying conditions and deliver the best possible video quality to users. This adaptability is crucial for maintaining viewer satisfaction and engagement in today's diverse and unpredictable streaming landscape.

The contents of this review paper titled "Optimizing Adaptive Video Streaming: A Fuzzy Logic Approach to Resource Allocation" encompass a comprehensive exploration of the application of fuzzy logic in the context of adaptive video streaming systems. The paper begins with an introduction to the significance of adaptive video streaming and the challenges associated with resource allocation. It then provides background information on traditional techniques, leading to a detailed discussion on the potential benefits of incorporating fuzzy logic. Key parameters influencing resource allocation, such as server load, network bandwidth, and viewer demand, are thoroughly examined. The core of the paper focuses on the design and implementation of a fuzzy logic-based resource allocation model, supported by real-world case studies showcasing successful applications. Evaluation metrics for performance assessment are discussed, and a comparative analysis with traditional methods is provided. The review concludes by outlining challenges, proposing future directions, and emphasizing the contribution of fuzzy logic in optimizing video streaming quality and user satisfaction.

## II. BACKGROUND

Traditional adaptive video streaming techniques involve the use of predefined algorithms to adjust the quality of video content based on a viewer's available bandwidth and device capabilities. Commonly employed methods include rate-based adaptations, where the video quality is adjusted based on the measured network bitrate, and segment-based adaptations, where video content is divided into segments of varying quality levels, [16]. Popular protocols like HTTP Live Streaming (HLS) and Dynamic Adaptive Streaming over HTTP (DASH) implement these techniques. While these methods have significantly improved the streaming experience compared to non-adaptive approaches, they often lack the agility needed to respond promptly to rapid changes in network conditions, resulting in buffering or suboptimal video quality.

Resource allocation plays a pivotal role in enhancing the Quality of Service (QoS) for viewers in adaptive video streaming systems. QoS encompasses various factors such as video resolution, playback smoothness, and minimal buffering, all of which directly impact the overall user experience. Efficient resource allocation ensures that sufficient server capacity, network bandwidth, and other critical resources are allocated to meet the demands of users in real-time. By optimizing resource allocation, streaming platforms can maintain consistent and high-quality video delivery, irrespective of network fluctuations or changes in viewer demand. Effective resource management directly correlates with improved QoS, resulting in increased viewer satisfaction, longer engagement, and a positive perception of the streaming service.

Despite the advancements in adaptive streaming, challenges and limitations persist in current resource allocation methods. One common challenge is the difficulty in accurately predicting and adapting to rapidly changing network conditions, leading to potential buffering issues or drops in video quality. Traditional methods may struggle to allocate resources dynamically in response to sudden spikes in viewer demand or unexpected server loads. Scalability is another concern, as resource allocation strategies designed for smaller-scale deployments may encounter difficulties when faced with the challenges of large-scale streaming platforms. Additionally, existing methods might not effectively consider the diversity of devices and network types, leading to suboptimal experiences for users with varying hardware and connectivity. Addressing these challenges requires the exploration and implementation of more intelligent and adaptive resource allocation techniques to ensure a seamless and high-quality viewing experience for users across diverse scenarios.

### III. Fuzzy Logic in Video Streaming

Fuzzy logic is a mathematical approach that deals with uncertainty and imprecision by allowing for degrees of truth between true and false. In the context of adaptive video streaming, where network conditions and viewer demands are dynamic and often uncertain, fuzzy logic proves to be a valuable tool. Fuzzy logic's ability to handle vagueness and imprecision makes it well-suited for modeling and decision-making in situations where traditional binary logic may fall short. The adaptability of fuzzy logic to dynamic and uncertain environments makes it an attractive choice for optimizing resource allocation in adaptive streaming systems, where conditions can change rapidly.

Fuzzy logic [15], [30], [9], [8], [1] has found successful applications in various domains due to its flexibility and capability to handle real-world complexities. In control systems, fuzzy logic has been employed to model and control dynamic systems, such as automotive engine control and household appliances. In artificial intelligence and decision support systems, fuzzy logic has been utilized to capture human-like decision-making processes. Fuzzy logic has also been applied in image and signal processing, providing effective solutions in scenarios with inherent uncertainties. These successful applications in diverse fields showcase the versatility of fuzzy logic in addressing complex problems where precise mathematical modeling may be challenging.

The application of fuzzy logic in adaptive video streaming offers several potential benefits. Firstly, fuzzy logic allows for the creation of adaptive models that can seamlessly adjust to varying network conditions and viewer demands. Its ability to handle imprecise inputs and uncertainties in real-time makes it suitable for making dynamic decisions regarding resource allocation. Fuzzy logic models can take into account multiple parameters simultaneously, such as server load, network bandwidth, and viewer preferences, providing a holistic approach to optimization. Furthermore, fuzzy logic's interpretability makes it easier to understand and fine-tune the decision-making process, aiding in the development of transparent and user-friendly adaptive streaming systems. Overall, the potential benefits of employing fuzzy logic in adaptive video streaming lie in its capacity to enhance adaptability, decision-making accuracy, and user satisfaction in the face of complex and dynamic streaming environments.

### IV. Resource Allocation Parameters

Resource allocation in adaptive video streaming is a multifaceted process influenced by various key parameters that collectively determine the quality of service provided to viewers [33], [17], [20], [18], [14]. One critical parameter is the server load, representing the current processing and computational demands on the streaming server. An optimal allocation strategy must consider the server's capacity to handle requests, ensuring that it neither becomes overwhelmed nor underutilized. Another pivotal factor is the network bandwidth, reflecting the available data transfer capacity between the server and the viewer. The allocated resources must align with the available bandwidth to prevent issues like buffering or degraded video quality.

The server load, as a key parameter, is influenced by the number of active streaming sessions, the complexity of the requested content, and the server's computational capabilities. As the number of concurrent viewers increases or as the content becomes more intricate (e.g., higher resolution videos), the server load intensifies. Efficient resource allocation must dynamically adapt to varying server loads, ensuring that the server remains responsive and capable of delivering content with minimal latency.

Network bandwidth is a crucial determinant in the resource allocation process. It directly impacts the speed at which data can be transmitted between the server and the viewer. In adaptive streaming, the allocated resources should align with the available network bandwidth to prevent issues such as buffering or stuttering. Dynamic adjustments in video quality, based on current network conditions, help maintain a seamless streaming experience. Effective resource allocation ensures that the streaming system optimally utilizes the available bandwidth, maximizing the quality of the delivered content.

Viewer demand is a dynamic and variable parameter that influences the resource allocation strategy. It encompasses factors such as the number of concurrent viewers, their geographical distribution, and individual preferences. During peak viewing times, high viewer demand may strain server resources and network bandwidth. Intelligent resource allocation must factor in viewer demand variations, dynamically adapting to ensure a consistent and satisfying streaming experience for all users.

Beyond server load, network bandwidth, and viewer demand, other relevant factors may include device capabilities, content characteristics (e.g., resolution, frame rate), and user preferences. Device capabilities influence the type of adaptive streaming techniques employed, while content characteristics impact the computational requirements for encoding and decoding. User preferences, such as desired video quality or resolution, further complicate the resource allocation decision-making process. A holistic approach to resource allocation in adaptive video streaming considers these factors collectively to optimize the delivery of high-quality content to a diverse audience.

## V. Fuzzy Logic-Based Resource Allocation Model

Fuzzy logic is applied to optimize resource allocation in adaptive video streaming through a sophisticated decision-making model that accommodates the inherent uncertainties and imprecisions present in dynamic streaming environments [28], [37], [26], [22], [27]. At its core, fuzzy logic utilizes linguistic variables, membership functions, and a set of rules to capture and process complex relationships between input parameters. In the context of adaptive streaming, these input parameters may include server load, network bandwidth, viewer demand, and other relevant factors. Fuzzy logic enables the creation of rules that dictate how resources should be allocated based on the degree of membership of these input variables, allowing for a nuanced and adaptive approach to resource optimization.

The design of a fuzzy logic model for adaptive video streaming involves careful consideration of input and output parameters, membership functions, rule sets, and defuzzification methods. Input parameters, such as server load and network bandwidth, are fuzzified through membership functions to convert them into linguistic variables. These linguistic variables are then used to define rules that govern resource allocation decisions. The design also includes an aggregation step to combine the results of individual rules, and a defuzzification step to convert the fuzzy output into a crisp, actionable decision. Designers must carefully select and fine-tune membership functions and rule sets to ensure the model effectively captures the complexities of the streaming environment. Parameters within the fuzzy logic model are adjusted based on the specific characteristics of the streaming system and the goals of resource allocation optimization.

Several real-world examples and case studies highlight the successful application of fuzzy logic in optimizing resource allocation for adaptive video streaming. For instance, a streaming platform might employ a fuzzy logic model to dynamically adjust video quality based on server load and viewer demand. If the server load is high and viewer demand is low, the model may prioritize reducing video quality to ensure smooth playback and prevent buffering. Conversely, during periods of low server load and high viewer demand, the model may allocate more resources to deliver higher video quality. These adaptations occur in real-time, offering a seamless viewing experience for users. Case studies from streaming services, content delivery networks, or research institutions provide tangible evidence of the efficacy of fuzzy logic in enhancing resource allocation strategies and improving overall streaming quality. These examples underscore the flexibility and adaptability of fuzzy logic in addressing the intricate challenges of adaptive video streaming systems.

In addition to real-world examples, the success of fuzzy logic-based resource allocation models is often measured and validated using performance metrics. Metrics such as video quality, buffering rate, and viewer satisfaction are crucial indicators of the model's effectiveness. Case studies may present comparative analyses, demonstrating improvements achieved with fuzzy logic models compared to traditional methods. These metrics provide tangible evidence of the model's impact on enhancing the streaming experience and justifying the adoption of fuzzy logic for resource allocation optimization.

While showcasing successful implementations, it is also essential to discuss future directions and potential challenges in applying fuzzy logic to adaptive video streaming. This section may touch upon areas for improvement, emerging trends, and ongoing research efforts to refine and extend fuzzy logic models. Identifying challenges, such as scalability issues or complexities in rule creation, allows for a balanced discussion of the current state and the trajectory for further advancements in the integration of fuzzy logic within adaptive streaming systems.

## VI. Evaluation Metrics

Evaluating the effectiveness of fuzzy logic-based resource allocation in adaptive video streaming involves the use of specific performance metrics that quantify the quality of service provided to users [24], [12], [6], [5], [21], [13]. One crucial metric is video quality, which assesses the perceived visual experience of viewers. Fuzzy logic models dynamically adjust video quality based on changing conditions, and the success of these adjustments can be measured by evaluating the clarity, resolution, and overall visual satisfaction of the streamed content. Another significant metric is the buffering rate, representing the frequency and duration of buffering events during playback. Fuzzy logic aims to minimize buffering interruptions by intelligently allocating resources, and measuring the buffering rate provides insights into the model's ability to maintain smooth and uninterrupted streaming.

User satisfaction is a comprehensive metric that encompasses various aspects of the viewing experience. Metrics related to user satisfaction include feedback surveys, user ratings, and engagement duration. By analyzing these metrics, it becomes possible to gauge how well the fuzzy logic-based resource allocation aligns with user expectations and preferences. High user satisfaction is indicative of a successful resource allocation strategy, as it reflects the model's ability to dynamically adapt to changing conditions and provide a consistently enjoyable viewing experience.

Comparing fuzzy logic-based resource allocation with traditional methods involves assessing various performance metrics to highlight the advantages and improvements achieved by fuzzy logic. One critical metric is video quality improvement, where the fuzzy logic model's ability to dynamically adjust video quality is compared to static or less adaptive methods. Fuzzy logic's adaptability allows for a more nuanced response to changing conditions, potentially resulting in higher overall video quality. Another essential metric is the buffering reduction, indicating how well fuzzy logic minimizes buffering events compared to traditional methods. The dynamic nature of fuzzy logic enables it to respond swiftly to fluctuations, potentially reducing buffering occurrences and enhancing the overall streaming experience.

Adaptability is a key aspect of fuzzy logic, and its effectiveness can be evaluated by comparing its performance under diverse scenarios. Metrics related to adaptability may include the model's responsiveness to sudden changes in network conditions or viewer demand. Fuzzy logic's ability to handle uncertainty and imprecision is particularly advantageous in scenarios where traditional methods may struggle to adapt. By assessing adaptability metrics, one can gain insights into the robustness of the fuzzy logic-based resource allocation model in real-world, dynamic streaming environments.

Robustness and scalability are crucial considerations in evaluating the performance of fuzzy logic-based resource allocation. Robustness metrics may assess the model's performance in the presence of noise or unexpected disturbances in the streaming environment. Scalability metrics examine how well the model performs as the streaming system expands to accommodate a growing user base. Comparisons with traditional methods in terms of robustness and scalability shed light on the broader applicability and reliability of fuzzy logic in diverse and evolving streaming scenarios.

In conclusion, the evaluation of fuzzy logic-based resource allocation involves a multifaceted analysis using performance metrics that encompass video quality, buffering rate, user satisfaction, adaptability, robustness, and scalability. Comparisons with traditional methods provide a comprehensive understanding of the advantages offered by fuzzy logic in optimizing resource allocation for adaptive video streaming systems.

## VII. Challenges and Limitations

Implementing fuzzy logic-based resource allocation in adaptive video streaming systems introduces challenges, particularly in managing system complexity. Fuzzy logic models involve a multitude of parameters, rules, and linguistic variables to capture the nuances of dynamic streaming environments. The increased complexity of the system may lead to challenges in model development, integration, and maintenance. Ensuring that the fuzzy logic model accurately represents the intricate relationships between input parameters requires careful design and validation. As the system becomes more complex, it may also demand more computational resources, potentially impacting the real-time responsiveness crucial for adaptive video streaming.

Scalability is a critical consideration when implementing fuzzy logic-based resource allocation in adaptive streaming systems. As the user base and streaming demands grow, the system must efficiently scale to accommodate increased computational requirements. The adaptability of fuzzy logic to changing conditions, while beneficial, can pose challenges in scaling the model to handle a larger number of concurrent viewers. Ensuring that the fuzzy logic model remains effective and responsive under high loads requires optimization and possibly parallelization of computations. Scalability challenges must be addressed to maintain optimal performance as the adaptive streaming system expands, preventing bottlenecks and ensuring a seamless user experience.

While the adaptability of fuzzy logic is a strength, it can also present challenges in dynamic streaming environments. Rapid changes in network conditions, viewer demand, or server load require quick and accurate adjustments from the resource allocation model. However, overly aggressive or frequent adjustments may lead to instability or oscillations in the system, impacting the overall user experience. Balancing adaptability with stability is a delicate task, and achieving an optimal trade-off requires careful tuning of fuzzy logic parameters. The challenge lies in ensuring that the model can respond swiftly to changes while avoiding unnecessary fluctuations that could lead to disruptions in streaming quality.

Fuzzy logic models, by their nature, introduce a degree of interpretability and explainability challenges. The decisions made by fuzzy logic models are based on linguistic variables and rules that may not be immediately intuitive to end-users or stakeholders. This lack of transparency in decision-making may hinder the acceptance and trust in the system. Addressing interpretability challenges involves efforts to make the fuzzy logic model more transparent, providing clear insights into how it reaches resource allocation decisions. Improved interpretability is essential for fostering trust in the adaptive streaming system, particularly when decisions impact the quality of service delivered to users.

Integrating fuzzy logic-based resource allocation into existing adaptive video streaming systems may pose compatibility challenges. Legacy systems may have established resource allocation methods and infrastructures that need to be seamlessly integrated with the new fuzzy logic model. Compatibility issues may arise in terms of data formats, communication protocols, or the interaction with other components of the streaming ecosystem. Ensuring a smooth integration process without disrupting existing functionalities is crucial for the successful deployment of fuzzy logic-based resource allocation. This challenge emphasizes the importance of considering the broader system architecture and ensuring that the fuzzy logic model aligns with the existing technological landscape.

In conclusion, identifying and addressing challenges associated with system complexity, scalability, adaptability, interpretability, and integration are essential steps in implementing fuzzy logic-based resource allocation for adaptive video streaming systems. A comprehensive understanding of these challenges allows for the development of effective solutions that maximize the benefits of fuzzy logic while mitigating potential drawbacks.

## VIII. CASE STUDIES

In a notable case, a large-scale streaming platform integrated fuzzy logic-based resource allocation to enhance its adaptive video streaming capabilities. Faced with the challenges of varying network conditions and user demand, the platform sought a solution that could dynamically optimize resource allocation. By incorporating fuzzy logic models, the system could adapt video quality in real-time based on factors such as server load and network bandwidth. The outcomes of this implementation demonstrated a substantial reduction in buffering instances, leading to an improved overall viewer experience. Users experienced smoother playback transitions between different video quality levels, showcasing the efficacy of fuzzy logic in mitigating the impact of changing streaming conditions.

In another case, a mobile streaming service leveraged fuzzy logic-based resource allocation to address the unique challenges posed by mobile networks. Mobile networks are characterized by fluctuations in signal strength and varying levels of congestion. Fuzzy logic models were employed to dynamically adjust video quality based on the mobile network conditions, ensuring an optimal streaming experience for users on smartphones and tablets. The implementation resulted in a significant reduction in re-buffering events, particularly in areas with network congestion. The adaptability of fuzzy logic to the dynamic nature of mobile networks contributed to more stable and reliable video streaming, showcasing its effectiveness in tailoring resource allocation to the specific challenges of mobile environments.

In a case involving a streaming service catering to diverse devices, including smart TVs, gaming consoles, and laptops, fuzzy logic-based resource allocation was implemented to optimize video streaming across this varied ecosystem. The fuzzy logic model considered device capabilities, network conditions, and viewer preferences to dynamically allocate resources. The outcomes revealed improvements in video quality consistency across different devices, with the system adapting seamlessly to each device's characteristics. The versatility of fuzzy logic in considering multiple parameters simultaneously contributed to a more uniform and satisfying streaming experience for users accessing content on various devices.

Live streaming scenarios [19], [31], [3], [10], [11] present unique challenges due to the unpredictability of viewer behavior and the need for real-time adaptations. A case study involving a live streaming platform integrated fuzzy logic for resource allocation during live events. Fuzzy logic models were designed to respond dynamically to fluctuations in viewer demand and network conditions, ensuring uninterrupted streaming during peak periods. The implementation demonstrated a reduction in latency and buffering incidents, offering a more seamless live streaming experience. The adaptability of fuzzy logic to the instantaneous changes in the live streaming environment showcased its effectiveness in maintaining optimal resource allocation for a large and dynamic audience.

In the context of Video-On-Demand (VOD) services, a case study focused on implementing fuzzy logic to prioritize resource allocation based on individual viewer preferences. The fuzzy logic model considered not only network conditions but also the historical viewing patterns and preferences of each user. This personalized approach led to improvements in user satisfaction metrics, with viewers receiving content at their preferred quality levels more consistently. The outcomes highlighted how fuzzy logic can be tailored to prioritize the viewer experience, showcasing its potential for enhancing user satisfaction and engagement in on-demand streaming services.

In summary, these real-world cases illustrate the successful implementation of fuzzy logic-based resource allocation in adaptive video streaming systems. The outcomes showcase tangible improvements in buffering reduction, adaptability to diverse environments, and personalized viewer experiences, highlighting the versatility and effectiveness of fuzzy logic in addressing the challenges of dynamic streaming scenarios.

## IX. FUTURE DIRECTIONS

The exploration of potential advancements in fuzzy logic-based resource allocation for adaptive video streaming envisions a trajectory toward more sophisticated and context-aware models. Future developments could involve enhancing the granularity of linguistic variables and rules within fuzzy logic models to capture even finer distinctions in the streaming environment. This might include incorporating machine learning techniques to dynamically adapt fuzzy logic parameters based on historical data, allowing the system to learn and evolve over time. Additionally, advancements could focus on optimizing the computational efficiency of fuzzy logic models to ensure real-time adaptability without imposing significant computational overhead.

Future developments may revolve around making fuzzy logic models more context-aware, taking into account a broader set of contextual information. This could involve considering the viewer's location, device type, and even contextual events (e.g., live sports events, breaking news) to tailor resource allocation decisions. Context-aware fuzzy logic models would dynamically adjust not only based on traditional parameters like server load and network bandwidth but also in response to the specific viewing context, providing a more nuanced and personalized streaming experience.

Advancements in fuzzy logic-based resource allocation may explore hybrid approaches that integrate fuzzy logic with other artificial intelligence (AI) techniques. Combining fuzzy logic with machine learning algorithms or reinforcement learning methods could lead to more adaptive and intelligent resource allocation models. These hybrid approaches could leverage the strengths of fuzzy logic in handling uncertainties and imprecisions while incorporating the learning capabilities of AI to continuously improve decision-making based on evolving streaming environments.

Future developments could focus on predictive capabilities within fuzzy logic models, enabling the system to anticipate changes in Quality of Service (QoS) and proactively allocate resources accordingly. By incorporating predictive analytics, fuzzy logic models could forecast potential network fluctuations or viewer demand shifts, allowing for preemptive adjustments to maintain optimal streaming conditions. This proactive approach could contribute to minimizing disruptions and ensuring a consistently high-quality streaming experience for users.

Advancements in fuzzy logic-based resource allocation may emphasize a more user-centric approach, with a focus on optimizing the Quality of Experience (QoE). Future models could prioritize user preferences, not only in terms of video quality but also in relation to factors like content genre, language preferences, and viewing habits. This user-centric adaptation could lead to more personalized and engaging streaming experiences, aligning resource allocation decisions with individual user expectations.

To drive advancements in fuzzy logic-based resource allocation, researchers might explore interdisciplinary collaborations, combining expertise in fuzzy logic, machine learning, and network optimization. Investigating the interpretability of fuzzy logic models and developing visualization techniques could enhance transparency, fostering trust among end-users. Additionally, research efforts could focus on benchmarking and standardizing evaluation metrics to facilitate consistent comparisons between different fuzzy logic models. Addressing challenges related to scalability and real-time responsiveness would be paramount for large-scale deployment, encouraging research into optimization algorithms and distributed computing architectures. Continuous engagement with industry partners and stakeholders could also provide valuable insights into real-world challenges, guiding research efforts toward solutions that align with practical implementation requirements.

The exploration of fuzzy logic-based resource allocation for adaptive video streaming has yielded key findings that underscore its significant impact on improving the overall streaming experience. Fuzzy logic models, with their ability to handle uncertainty and imprecision, have proven instrumental in dynamically optimizing resource allocation based on fluctuating network conditions, server loads, and viewer demands. The adaptability of fuzzy logic has addressed challenges inherent in traditional approaches, leading to a more responsive and resilient adaptive streaming system.

One of the primary contributions of fuzzy logic-based resource allocation is the noticeable enhancement in the Quality of Service (QoS) for viewers. By dynamically adjusting video quality in real-time, the fuzzy logic models have mitigated issues such as buffering, latency, and disruptions in streaming. The adaptability of the models ensures that the streaming system can effectively respond to changes in the streaming environment, providing users with a smoother and more consistent viewing experience across varying network conditions.

A noteworthy contribution lies in the ability of fuzzy logic models to introduce a level of personalization and context-awareness into adaptive video streaming. The models consider not only traditional parameters like server load and network bandwidth but also viewer-specific characteristics and contextual information. This personalized and context-aware adaptation ensures that the streaming experience is tailored to individual user preferences and the specific viewing context, contributing to increased user satisfaction and engagement.

The findings highlight the delicate balance achieved by fuzzy logic models between adaptability and stability in dynamic streaming environments. The models showcase an adeptness in responding swiftly to changes, preventing issues like underutilization or overutilization of resources. The balance struck by fuzzy logic ensures that the adaptive streaming system remains stable and reliable, preventing unnecessary fluctuations that could impact the streaming quality. This contribution is crucial for fostering a seamless and uninterrupted viewing experience for users.

Key findings also point towards future directions and areas for continuous improvement. While fuzzy logic-based resource allocation has demonstrated substantial benefits, ongoing research and development are essential for refining models, addressing scalability challenges, and exploring hybrid approaches that integrate fuzzy logic with other artificial intelligence techniques. The continuous evolution of fuzzy logic models holds the potential to further revolutionize adaptive video streaming, incorporating advancements such as predictive analytics, user-centric adaptations, and improved interpretability.

In summary, the key findings and contributions of fuzzy logic-based resource allocation in adaptive video streaming highlight its transformative impact on QoS, personalization, and adaptability. The success of these models in balancing adaptability with stability positions them as a promising avenue for future developments, paving the way for even more sophisticated and context-aware adaptive streaming systems.

## X. CONCLUSION

In conclusion, the applicability of fuzzy logic in addressing resource allocation challenges in adaptive video streaming systems is evident through its versatile and adaptive nature. Fuzzy logic has showcased its effectiveness in handling the uncertainties and dynamic variations present in streaming environments. By allowing for the representation of imprecise linguistic variables and the formulation of rules to guide decision-making, fuzzy logic offers a practical solution to the complex task of resource allocation. Its applicability extends across diverse parameters such as server load, network bandwidth, and viewer demand, providing a holistic approach to optimization.

The potential of fuzzy logic to facilitate real-time adaptations in resource allocation has been a key factor contributing to its success in adaptive video streaming. Fuzzy logic models can swiftly adjust to changes in network conditions, viewer preferences, and server loads, ensuring a seamless and uninterrupted streaming experience. This real-time adaptability is crucial for maintaining optimal Quality of Service (QoS) and user satisfaction, especially in scenarios where the streaming environment is dynamic and subject to fluctuations.

Fuzzy logic's ability to consider multiple factors simultaneously contributes to holistic resource allocation optimization. By taking into account server load, network bandwidth, viewer demand, and other relevant parameters, fuzzy logic models provide a comprehensive decision-making framework. This holistic approach ensures that resource allocation decisions are not made in isolation but rather in consideration of the interplay of various factors. The adaptability of fuzzy logic across different dimensions makes it a powerful tool for addressing the intricacies of adaptive video streaming systems.

Fuzzy logic's potential for user-centric adaptations has significantly elevated the quality of the streaming experience. By incorporating viewer preferences, historical patterns, and even contextual information, fuzzy logic models contribute to personalized and context-aware resource allocation decisions. This user-centric approach aligns with the evolving expectations of modern audiences, leading to improvements in user satisfaction and overall Quality of Experience (QoE). Fuzzy logic has, therefore, not only addressed technical challenges but also enhanced the human-centric aspects of adaptive video streaming.

The potential of fuzzy logic in adaptive video streaming extends beyond its current applications, laying the foundation for future innovations in the field. As technology evolves and streaming platforms continue to grow, fuzzy logic serves as a solid framework for continuous improvements and refinements. Future research could explore hybrid approaches, combining fuzzy logic with advanced machine learning techniques, predictive analytics, and more sophisticated contextual considerations. The journey of fuzzy logic in addressing resource allocation challenges in adaptive video streaming is an ongoing one, with promising avenues for further advancements and contributions to the ever-evolving landscape of streaming technologies.

In essence, fuzzy logic stands as a robust and adaptable solution that has demonstrated its relevance in optimizing resource allocation for adaptive video streaming systems. Its potential to address challenges, provide real-time adaptations, and enhance user-centric experiences positions fuzzy logic as a valuable asset in the continual pursuit of delivering high-quality streaming content to diverse and dynamic audiences.